\documentclass[journal]{IEEEtran}

\usepackage{booktabs}
\usepackage[dvipsnames]{xcolor}

\usepackage{amsmath}

\usepackage[linesnumbered,ruled,vlined]{algorithm2e}
\SetAlTitleFnt{\footnotesize}
\SetAlCapNameFnt{\footnotesize}
\SetAlCapFnt{\footnotesize}
\usepackage{adjustbox}
\usepackage{soul,xcolor}
\usepackage{graphicx,dblfloatfix}
\usepackage{multirow}
\usepackage{rotating}
\usepackage{array}
\usepackage{enumitem}
\usepackage[normalem]{ulem}
\usepackage[font=small,labelfont=bf,skip=3pt]{caption}
\usepackage{ragged2e}
\usepackage[utf8]{inputenc}
\usepackage[T1]{fontenc}
\usepackage{pifont}
\usepackage{amssymb}
\usepackage{stix}
\usepackage{bm}
\usepackage[numbers,square,comma,compress,sort]{natbib}

\usepackage{tikz}

\setstcolor{red}
\newcommand{\new}[1]{\textcolor{black}{#1}}

\begin{document}
\title{Runtime Task Scheduling using Imitation Learning for Heterogeneous Many-Core Systems}


\author{
Anish Krishnakumar, Samet E. Arda, A. Alper Goksoy, Sumit K. Mandal, \\ Umit Y. Ogras, Anderson L. Sartor, Radu Marculescu
\thanks{
Manuscript received April 17, 2020; revised June 12, 2020; accepted July 6, 2020. Date of current version July 16, 2020.
This article was presented in the International Conference on Hardware/Software Codesign and System Synthesis 2020 and appears as part of the ESWEEK-TCAD special issue. 
\textit{(Corresponding author: Anish Krishnakumar.)} 
\newline \indent This material is based on research sponsored by Air Force Research Laboratory (AFRL) and Defense Advanced Research Projects Agency (DARPA) under agreement number FA8650-18-2-7860. The U.S. Government is authorized to reproduce and distribute reprints for Governmental purposes notwithstanding any copyright notation thereon. The views and conclusion contained herein are those of the authors and should not be interpreted as necessarily representing the official policies or endorsements, either expressed or implied, of Air Force Research Laboratory (AFRL) and Defense Advanced Research Projects Agency (DARPA) or the U.S. Government.\newline
\indent A. Krishnakumar, S. E. Arda, A. A. Goksoy, S.K. Mandal and U. Y. Ogras are with the School of Electrical, Computer and Energy Engineering, Arizona State University, Tempe, AZ 85287 USA. \newline E-mail: \{anishnk, sarda1, aagoksoy, skmandal, umit\}@asu.edu \newline
\indent A. L. Sartor is with the Electrical and Computer Engineering Dept., Carnegie Mellon University, Pittsburgh, PA 15213 USA. \newline E-mail: \{asartor\}@cmu.edu \newline
\indent R. Marculescu is with the Electrical and Computer Engineering Dept., The University of Texas at Austin, Austin, TX 78712 USA. \newline
E-mail: \{radum\}@utexas.edu
\newline \indent
Digital Object Identifier 10.1109/TCAD.2020.3012861
\newline
© 2020 IEEE. Personal use of this material is permitted. Permission from IEEE must be obtained for all other uses, in any current or future media, including reprinting/republishing this material for advertising or promotional purposes, creating new collective works, for resale or redistribution to servers or lists, or reuse of any copyrighted component of this work in other works.
}
}




\IEEEtitleabstractindextext{
\begin{abstract}
\label{sec:abstract}
Domain-specific systems-on-chip, a class of heterogeneous many-core systems, are recognized as a key approach to narrow down the performance and energy-efficiency gap 
between custom hardware accelerators and programmable processors.
Reaching the full potential of these architectures 
depends critically on optimally scheduling the applications 
to available resources at runtime. 
Existing optimization-based techniques cannot achieve this objective at runtime due to the combinatorial nature of the task scheduling problem. 
As the main theoretical contribution, this paper poses scheduling as a classification problem 
and proposes a hierarchical imitation learning (IL)-based scheduler that learns from an Oracle to maximize the performance of multiple domain-specific applications. 
Extensive evaluations with six streaming applications 
from wireless communications and radar domains 
show that the proposed IL-based scheduler
approximates an offline Oracle policy with more than 99\% accuracy \new{for performance- and energy-based optimization objectives.} 
Furthermore, it achieves almost identical performance to the Oracle with a low runtime overhead
and successfully adapts to new applications, \new{many-core system configurations, and runtime variations in application characteristics}.
\end{abstract}

}

\maketitle
\IEEEdisplaynontitleabstractindextext
\hspace{2mm}
\section{Introduction} 
\label{sec:introduction}

\IEEEPARstart{H}{omogeneous} multi-core architectures have successfully exploited thread- and data-level parallelism to achieve performance and energy efficiency beyond the limits of single-core processors. 
While general-purpose computing achieves programming flexibility, it suffers from significant performance and energy efficiency gap when compared to special-purpose solutions.
Domain-specific architectures, such as graphics processing units (GPUs) and neural network processors, are recognized as some of the most promising solutions to reduce this gap~\cite{hennessy2019new}. Domain-specific systems-on-chip (DSSoCs), a concrete instance of this new architecture, judiciously combine general-purpose cores, special-purpose processors, and hardware accelerators.
DSSoCs approach the efficacy of fixed-function solutions
for a specific domain while maintaining programming flexibility for other domains~\cite{green2018heterogeneous}.

The success of DSSoCs depends critically on satisfying two intertwined requirements. 
First, the available processing elements (PEs) must be utilized optimally, at runtime, to execute the incoming tasks. 
For instance, scheduling all tasks to general-purpose cores may work, but diminishes the benefits of the special-purpose PEs. 
Likewise, a static task-to-PE mapping could unnecessarily stall the parallel instances of the same task. 
Second, acceleration of the domain-specific applications needs to be oblivious to the application developers to make DSSoCs practical. 
This paper addresses these two requirements simultaneously. 

The task scheduling problem involves assigning tasks to processing elements and ordering their execution to achieve the optimization goals, e.g., minimizing execution time, power dissipation, or energy consumption. 
To this end, applications are abstracted using mathematical models, such as directed acyclic graph (DAG) and synchronous data graphs (SDG) that capture both the attributes of individual tasks (e.g., expected execution time) and the dependencies among the tasks~\cite{topcuoglu2002performance, baskiyar2010energy, sakellariou2004hybrid}. 
Scheduling these tasks to the available PEs is a well-known NP-complete problem~\cite{garey1979Computers, ullman1975npcomplete}. 
An optimal \textit{static schedule} can be found for small problem sizes using optimization techniques, such as mixed-integer programming (MIP)~\cite{goel2015constraint} and constraint programming (CP)~\cite{rossi2006handbook}. 
These approaches are not applicable to runtime scheduling for two fundamental reasons. 
First, statically computed schedules lose relevance in a dynamic environment where tasks from multiple applications stream in parallel, and PE utilizations change dynamically. 
Second, the execution time of these algorithms, hence their overhead, can be prohibitive even for small problem sizes with few tens of tasks. 
Therefore, a variety of heuristic schedulers, such as shortest job first (SJF)~\cite{vasile2015resource} and complete fair schedulers (CFS)~\cite{pabla2009completely}, are used in practice for homogeneous systems.
These algorithms trade off the quality of scheduling decisions and computational overhead.

To improve this state of affairs, this paper addresses the following challenging proposition: \textit{Can we achieve a scheduler performance close to that of optimal MIP and CP schedulers, while using minimal runtime overhead compared to commonly used heuristics?} 
Furthermore, we investigate this problem in the context of heterogeneous PEs.
We note that 
much of the scheduling in heterogeneous many-core systems is 
tuned manually, even to date~\cite{aji2016multicl}. 
For example, OpenCL, a widely-used programming model for heterogeneous cores, leaves the scheduling problem to the programmers.
Experts manually optimize the task to resource mapping based on their knowledge of application(s), characteristics of the heterogeneous clusters, data transfer costs, and platform architecture.
However, manual optimization suffers from scalability for two reasons.  First, optimizations do not scale well for all applications. 
Second, extensive engineering efforts are required to 
adapt the solutions to different platform architectures 
and varying levels of concurrency in applications.
Hence, there is a critical need for a methodology to provide optimized scheduling solutions applicable to a variety of applications at runtime in heterogeneous many-core systems.

Scheduling has traditionally been considered as an optimization problem~\cite{goel2015constraint}.
We change this perspective by formulating 
runtime scheduling for heterogeneous many-core platforms as a classification problem. 
This perspective and the following \textit{key insights} enable us to employ machine learning (ML) techniques to solve this problem:

\noindent \textit{Key insight 1:} 
One can use an optimal (or near-optimal) scheduling algorithm offline without being limited by computational time and other runtime overheads. 
Then, the inputs to this scheduler and its decisions can be recorded along with relevant features to construct an Oracle.

\noindent \textit{Key insight 2:} 
One can design a policy that approximates the Oracle with minimum overhead and use this policy at runtime.

\noindent \textit{Key insight 3:} 
One can exploit the effectiveness of ML to learn from Oracle with different objectives, which includes minimizing execution time, energy consumption, etc.

Realizing this vision requires addressing several challenges.
First, we need to construct an Oracle in a dynamic environment
where tasks from multiple applications can overlap arbitrarily, and each incoming application instance observes a different system state.
Finding optimal schedules is challenging even offline, since the underlying problem is NP-complete. 
We address this challenge by constructing Oracles using both CP
and a computationally expensive heuristic, called earliest task first (ETF)~\cite{hwang1989scheduling}.
ML uses informative properties of the system (\textit{features}) to predict the category in a classification problem. 
The second challenge is identifying the minimal set of relevant features that can lead to high accuracy with 
minimal overhead.
We store a small set of 45 relevant features for a many-core platform with 16 processing elements along with the Oracle to minimize the runtime overhead.
This enables us to represent a complex scheduling decision as a set of features and then predict the best processing element for task execution.
The final challenge is approximating the Oracle accurately with a minimum implementation overhead.
Since runtime task scheduling is a sequential decision-making problem, 
supervised learning methodologies, such as linear regression and regression tree, may not generalize for unseen states at runtime. 
Reinforcement learning (RL) and imitation learning (IL) 
are more effective for sequential decision-making problems~\cite{sutton2000policy, mandal2019dynamic, sartor2020hilite}. 
Indeed, RL has shown promise when applied to the scheduling problem~\cite{mao2016resource,mao2019learning,wang2019multi}, but it suffers from slow convergence and sensitivity of the reward function~\cite{kim2017imitation,mandal2020energy}.
In contrast, IL takes advantage of the expert's inherent knowledge and produces policies that imitate the expert decisions~\cite{schaal1999imitation}.
Hence, we propose an IL-based framework
that schedules incoming applications 
to heterogeneous multi-core systems.

The proposed IL framework is formulated to facilitate generalization, i.e. it can be adapted to learn from any Oracle
that optimizes a specific objective, such as performance and energy efficiency, of an arbitrary DSSoC.
We evaluate the proposed framework with six domain-specific applications from wireless communications and radar systems.
The proposed IL policies successfully approximate the Oracle with more than 99\% accuracy, achieving fast convergence and generalizing to unseen applications. 
In addition, the scheduling decisions are made within 1.1$\mu$s (on an Arm A53 core), which is better than CFS performance~(1.2$\mu$s).
To the best of our knowledge, this is the first imitation learning-based scheduling framework for heterogeneous many-core systems capable of handling multiple applications exhibiting streaming behavior.
The main contributions of this paper are as follows:
\begin{itemize} [leftmargin=*]
    \item An imitation learning framework to construct policies for task scheduling in heterogeneous many-core platforms;
    \item Oracle design using both optimal and heuristic schedulers for performance- and energy- based optimization objectives;
    \item Extensive experimental evaluation of the proposed IL policies along with latency and storage overhead analysis;
    \item Performance comparison of IL policies against reinforcement learning and optimal schedules obtained by constraint programming.
\end{itemize}

The rest of the paper is organized as follows.
We review the related work in Section~\ref{sec:related_work}. 
Section~\ref{sec:background} provides background information on DAG scheduling and imitation learning.
In Section~\ref{sec:methodology}, we discuss the proposed methodology, followed by relevant experimental results in Section~\ref{sec:experimental_results}.
Section~\ref{sec:conclusion} presents the conclusions and possible future research for this work.

\section{Related Work and Novel Contributions} 
\label{sec:related_work}

\begin{figure*}[t!]
	\centering
	\includegraphics[width=1.0\linewidth]{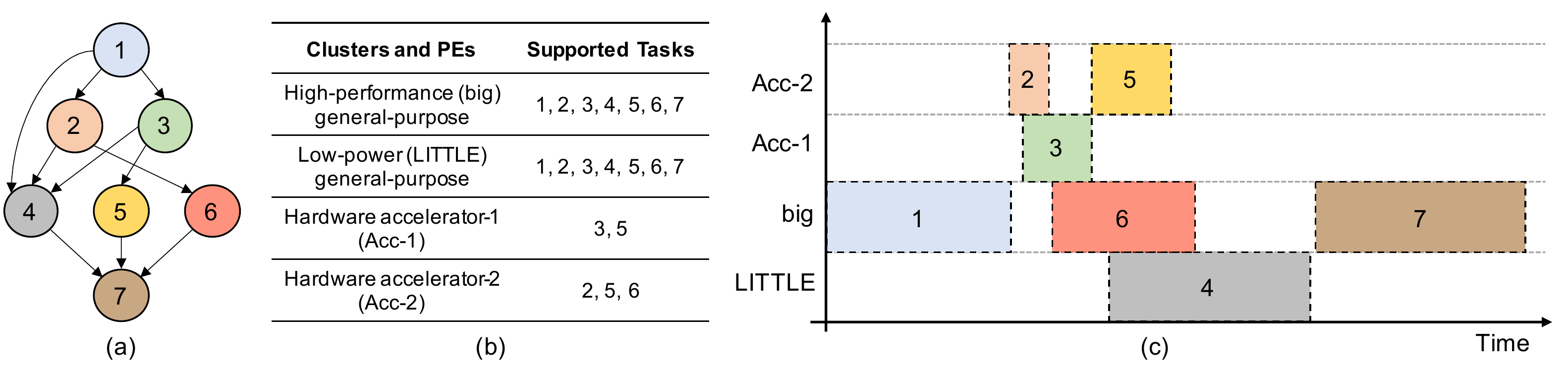}
	\caption{(a) An example DAG consisting of 7 tasks (b) A heterogeneous computing platform with 4 processing elements and list of tasks in DAG supported by each PE (c) A sample schedule of the DAG on the heterogeneous many-core system.}
	\label{fig:dag_schedule}
\end{figure*} 

Current many-core systems use runtime heuristics to enable scheduling with low overheads.
For example, the completely fair scheduler (CFS)~\cite{pabla2009completely}, widely used in Linux systems, aims to provide fairness for all processes in the system.
CFS maintains two queues (active and expired) to manage task scheduling. 
In addition, CFS gives a fixed time quantum for each process.
Tasks are swapped between active and expired queues based on activation and expiration of the time quantum.
However, complex heuristics are required to manage such queues.
CFS also does not generalize to optimization objectives apart from performance and fairness.
More importantly, CFS scheduling is limited to general-purpose cores and lacks support for specialized cores and hardware accelerators~\cite{beisel2011cooperative}.
With the same limitations, shortest job first (SJF)~\cite{vasile2015resource} scheduler estimates the task's CPU processing time and assigns the first available resource to the task with the shortest execution time.


List scheduling techniques~\cite{sakellariou2004hybrid, kwok1996dynamic} for DAGs~\cite{topcuoglu2002performance, bittencourt2010dag, arabnejad2013list} prioritize various objectives, such as energy~\cite{baskiyar2010energy, swaminathan2001real}, fairness~\cite{xie2016mixed}, security~\cite{xiaoyong2010novel}.
In general, this technique places the nodes (tasks) of a DAG in a list and provides a PE assignment and order at design time.
Heterogeneous earliest finish time (HEFT)~\cite{topcuoglu2002performance} is one example, in which an upward rank is computed to perform the scheduling decisions.
The authors in~\cite{bittencourt2010dag} use a lookahead algorithm as an enhancement to the HEFT scheduler to improve the execution time, but suffers from fourth order complexity $O(n^{4})$ on the number of tasks ($n$).
Another recent technique shows improvement in performance with quadratic complexity~\cite{arabnejad2013list}.
However, these algorithms suffer from the time complexity problem and are tailored to particular objectives
and fail to generalize to a combination of objectives and choice of applications.

\begin{figure*}[t]
	\centering
	\includegraphics[width=1.0\linewidth]{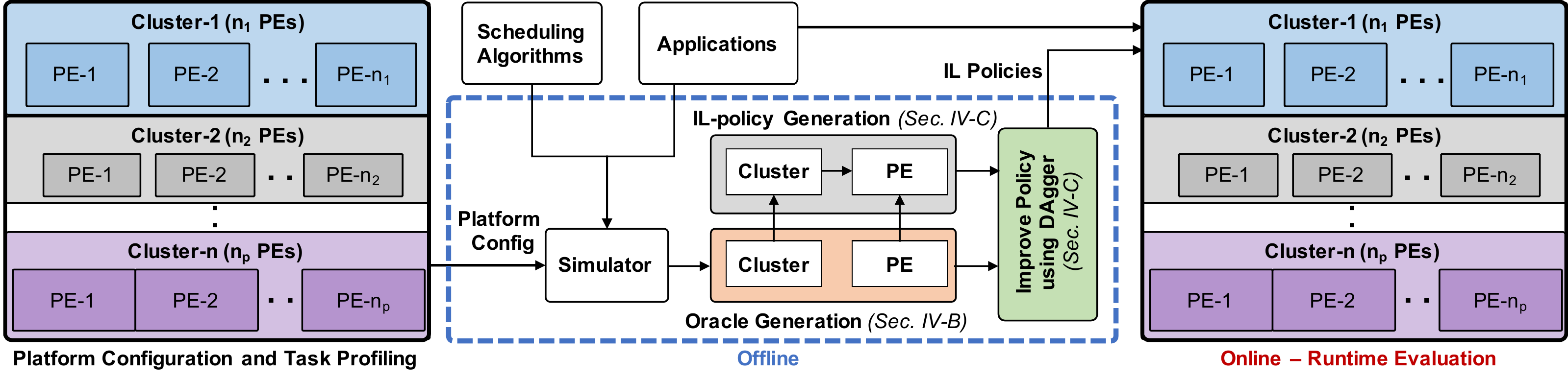}
	\caption{An overview of the proposed imitation learning framework for task scheduling in heterogeneous many-core systems. The framework integrates the system configuration, profiling information, scheduling algorithms and applications to construct Oracle, and train IL policies for task scheduling. The IL policies, that are improved using DAgger, are then evaluated on the heterogeneous many-core system at runtime.}
	\label{fig:framework}
\end{figure*}

Machine learning (ML)-based schedulers show promise in eliminating the drawbacks of list scheduling and runtime heuristic techniques.
ML-based schedulers possess the capabilities to be further tuned at runtime~\cite{mao2016resource}. 
A recent support vector machine (SVM)-based scheduler for OpenCL kernels assigns kernels (tasks) between CPUs and GPUs~\cite{wen2014smart}.
In contrast to schedulers that use supervised learning, authors in~\cite{mirhoseini2017device} uses reinforcement learning (RL) to schedule Tensorflow device placement, but lacks the ability of scheduling streaming jobs.
DeepRM~\cite{mao2016resource} uses deep neural networks with RL for scheduling at an application granularity as opposed to using the notion of DAGs.
On the other hand, Decima~\cite{mao2019learning} uses a combination of graph neural networks and RL to perform coarse-grained processing-cluster level scheduling for streaming DAGs.

RL-based scheduling techniques have two major drawbacks. 
\textit{First,} they require a significant number of episodes to converge. 
For example, the technique proposed in~\cite{mao2019learning} takes 50k episodes, with 1.5 seconds each, to converge to a solution that is equivalent to 21 hours of simulation in Nvidia Tesla P100 GPU.
\textit{Second}, the efficiency of an RL-based technique predominantly depends on the choice of the reward function. Usually, the reward function is hand-tuned, depending on the problem under consideration.

To overcome these difficulties, we propose an IL-based scheduling methodology.
Since IL uses an Oracle to construct a policy, it does not suffer from slow convergence, as seen in RL.
IL-based policies were initially used in robotics to show their fast convergence property~\cite{schaal1999imitation}.
Recently, the use of imitation learning to intelligently manage power and energy consumption in SoCs has been demonstrated~\cite{kim2017imitation, mandal2020energy}.
To the best of our knowledge, \textit{this is the first approach that applies IL for multi-application streaming task scheduling in heterogeneous many-core platforms}.

\section{Background and Overview}
\label{sec:background}

The runtime scheduling problem addressed in this paper is illustrated in Fig.~\ref{fig:dag_schedule}. 
We consider streaming applications that can be modeled using 
directed acyclic graphs, such as the one shown in Fig.~\ref{fig:dag_schedule}(a).
These applications process data frames that arrive at a varying rate over time.
For example, a WiFi-transmitter, one of our domain applications, receives and encodes raw data frames before they are transmitted over the air. 
Data frames from a single application or multiple simultaneous applications can overlap in time as they go through the tasks that compose the application. 
For instance, Task-1 in Fig.~\ref{fig:dag_schedule}(a) can start processing a new frame, while other tasks continue processing earlier frames. Processing of a frame is said to be completed after the terminal task without any successor (Task-7 in Fig.~\ref{fig:dag_schedule}(a)) is executed. 
We define the application formally to facilitate description of the schedulers.

\noindent\textbf{Definition 1:} An \textit{application graph} 
$G_{App}(\mathcal{T},\mathcal{E})$ 
is a directed acyclic graph, 
where each node $T_i \in \mathcal{T}$ 
represents the tasks that compose the application. 
Directed edge $e_{ij} \in \mathcal{E}$ from 
task $T_i$ to $T_j$ shows that $T_j$ cannot start processing a new frame before the output of $T_i$ reaches $T_j$ for all $T_i, T_j \in \mathcal{T}$. 
$v_{ij}$ for each edge $e_{ij} \in \mathcal{E}$ denotes the communication volume over this edge. 
It is used to account for the communication latency.

Each task in a given application graph $G_{App}$ 
can execute on different processing elements in the target DSSoC.
We formally define the target DSSoC as follows:

\noindent\textbf{Definition 2:} An \textit{architecture graph} 
$G_{Arch}(\mathcal{P},\mathcal{L})$ 
is a directed graph, 
where each node $P_i \in \mathcal{P}$ 
represents processing elements,
and $L_{ij} \in \mathcal{L}$ represents the communication links 
between $P_i$ and $P_j$ in the target SoC. 
The nodes and links have the following quantities associated with them:
\begin{itemize}
    \item $t_{exe}(P_i,T_j)$ is the execution time of task $T_j$ 
    on PE $P_i \in \mathcal{P}$, if $P_i$ can execute (i.e., it supports) $T_j$. 
    
    \item $t_{comm}(L_{ij})$ is the communication latency from 
    $P_i$ to $P_j$ for all $P_i, P_j \in \mathcal{P}$.
    
    \item $C(P_i) \in \mathcal{C}$ is the PE cluster $P_i \in \mathcal{P}$ belongs to.
\end{itemize}
The DSSoC example in Fig.~\ref{fig:dag_schedule}(b) 
assumes one big core cluster, one LITTLE core cluster, 
and two hardware accelerators each with a single PE in them 
for simplicity. 
The low-power (LITTLE) and high-performance (big) general-purpose clusters can support the execution of all tasks, as shown in the \textit{supported tasks} column in Fig.~\ref{fig:dag_schedule}(b).
In contrast, hardware accelerators 
(Acc-1 and Acc-2)
support only a subset of tasks.

\begin{table}[tb]
\centering
\caption{Summary of the notations used in this paper}
\label{tab:notations}
\setlength\tabcolsep{2pt}
\begin{tabular}{@{}ll|ll@{}}
\toprule
$T_{j}$ & Task-j & $\mathcal{T}$ & Set of Tasks \\ \midrule
$P_{i}$ & PE-i & $\mathcal{P}$ & Set of PEs \\ \midrule
$c$ & Cluster-$c$ & $\mathcal{C}$ & Set of clusters \\ \midrule
$L_{ij}$ & \begin{tabular}[c]{@{}l@{}}Communication links \\ between $P_i$ to $P_j$\end{tabular} & $\mathcal{L}$ & \begin{tabular}[c]{@{}l@{}}Set of \\ communication links\end{tabular} \\ \midrule
$t_{exe}(P_i,T_j)$ & \begin{tabular}[c]{@{}l@{}}Execution time of \\ task $T_{j}$ on PE $P_{i}$\end{tabular} & $t_{comm}(L_{ij})$ & \begin{tabular}[c]{@{}l@{}}Communication \\ latency from $P_i$ to $P_j$\end{tabular} \\ \midrule
$s$ & State-s & $\mathcal{S}$ & Set of states \\ \midrule
$v_{jk}$ & \begin{tabular}[c]{@{}l@{}}Communication volume \\ from task $T_{j}$ to $T_{k}$\end{tabular} & $\mathcal{A}$ & Set of actions \\ \midrule
$\mathcal{F}_{S}$ & Static features & $\mathcal{F}_{D}$ & Dynamic features \\ \midrule
$\pi_C(s)$ & \begin{tabular}[c]{@{}l@{}}Apply cluster policy\\ for state $s$\end{tabular} & $\pi_{P,c}(s)$ & \begin{tabular}[c]{@{}l@{}}Apply PE policy\\ in cluster-c for state $s$\end{tabular} \\ \midrule
$\pi$ & Policy & $\pi^{*}$ & Oracle policy \\ \midrule
\new{$\pi^{G}$} & \begin{tabular}[c]{@{}l@{}} \new{Policy for many-core} \\ \new{platform configuration G} \end{tabular} & \new{$\pi^{*G}$} & \begin{tabular}[c]{@{}l@{}} \new{Oracle for many-core} \\ \new{platform configuration G} \end{tabular} \\ \bottomrule
\end{tabular}
\end{table}


A particular instance of the scheduling problem is illustrated in Fig.~\ref{fig:dag_schedule}(c).
Task-6 is scheduled to big core (\textit{although it executes faster on 
Acc-2}) 
since 
Acc-2
is not available at the time of decision making.
Similarly, Task-4 is scheduled to the LITTLE core (\textit{even if it executes faster on big}) because the big core is utilized 
when Task-4 is ready to execute.
In general, scheduling complex DAGs in heterogeneous many-core platforms present a multitude of choices making the runtime scheduling problem highly complex.
The complexity increases further with: (1) overlapping DAGs at runtime, (2) executing multiple applications simultaneously,
and (3) optimizing for objectives such as performance, energy, etc.

The proposed solution leverages imitation learning, and is outlined in Fig.~\ref{fig:framework}.
It is also referred to as learning by demonstration, and is an adaption of supervised learning for sequential decision making problems.
The decision-making space is segmented into distinct decision epochs, called \textit{states}~($\mathcal{S}$).
There exists a finite set of actions $\mathcal{A}$ for every state $s \in \mathcal{S}$. 
IL uses policies that map each state ($s$) to a corresponding action.

\noindent\textbf{Definition 3: Oracle Policy (expert)} 
$\pi^*(s): \mathcal{S} \rightarrow \mathcal{A}$ 
maps a given system state to the optimal action. 
In our runtime scheduling problem, 
the state includes the set of ready tasks and actions that
correspond to assignment of tasks $\mathcal{T}$ to processing elements $\mathcal{P}$. 
Given the Oracle $\pi^*$, the goal with imitation learning is to learn a runtime policy that can approximate it. 
We construct an Oracle offline and approximate it using a hierarchical policy with two levels. 
Consider a generic heterogeneous many-core platform 
with a set of clusters $\mathcal{C}$, as illustrated in Fig.~\ref{fig:framework}. 
At the first level, an IL policy chooses one cluster (among $n$ clusters) for a task to be executed in.

The first-level policy assigns the ready tasks to one of the clusters in $\mathcal{C}$,  since each PE within the same cluster has the same static parameters. 
Then, a cluster-level policy assigns the tasks to a specific PE within that cluster. 
The details of state representation, Oracle generation, 
and hierarchical policy design are presented in the next section.

\section{Proposed Methodology and Approach} 
\label{sec:methodology}

This section first introduces the system state representation, including the features used by the IL policies. 
Then, it presents the Oracle generation process, and 
the design of the hierarchical IL policies.
Table~\ref{tab:notations} details the notations that will be used hereafter. 


\subsection{System State Representation}


Offline scheduling algorithms are NP-complete even though they rely on static features, such as average execution times.
The complexity of runtime decisions is further exacerbated as the system schedules multiple applications that exhibit streaming behavior.
In the streaming scenario, incoming frames do not observe an empty system with idle processors. 
In strong contrast, PEs have different utilization, and there may be an arbitrary number of partially processed frames in the wait queues of the PEs.
Since our goal is to learn a set of policies that generalize to all applications and all streaming intensities, the ability to learn the scheduling decisions critically depends on the effectiveness of state representation.
The system state should encompass both static and dynamic aspects of the set of tasks, applications, and the target platform. 
Naive representations of DAGs include adjacency matrix
and adjacency list.
However, these representations suffer from drawbacks such as large storage requirements, highly sparse matrices which complicates the training of supervised learning techniques, and scalability for multiple streaming applications.
In contrast, we carefully study the factors that influence task scheduling in a streaming scenario and construct features that accurately represent the system state.
We broadly categorize the features that make up the state as follows:
\begin{itemize}
    \item \textit{Task features:} This set includes the attributes of individual tasks. They can be both static, such as average execution time of a task on a given PE ($t_{exe}(P_i,T_j)$), and dynamic, such as the relative order of a task in the queue.    
    
    \item \textit{Application features:}
    This set describes the characteristics of the 
    entire application. They are static features, such as the number of tasks in the application and the precedence constraints between them.
    
    \item \textit{PE features:} This set describes the dynamic state of the processing elements. Examples include the earliest available times (readiness) of the PEs to execute tasks.
    
\end{itemize}
The static features are determined at the design time, whereas the dynamic features can only be computed at runtime.
The static features aid in exploiting design time behavior.
For example, 
$t_{exe}(P_i,T_j)$ 
helps the scheduler compare the expected performance of different PEs.
Dynamic features, on the other hand, present the runtime dependencies between tasks and jobs and also the busy states of the processing elements.
For example, the expected time when cluster $c$ becomes available for processing adds invaluable information, which is only available at runtime.

\begin{table}[t]
\centering
\caption{\new{Types of} features employed for state representation from point of view of task $T_{j}$}
\label{tab:feature_selection_2}
\begin{tabular}{@{}clc@{}}
\toprule
\textbf{\begin{tabular}[c]{@{}c@{}}Feature Type\end{tabular}} & \multicolumn{1}{c}{\textbf{\begin{tabular}[c]{@{}c@{}}Feature Description\end{tabular}}} & \textbf{\begin{tabular}[c]{@{}c@{}}Feature Categories\end{tabular}} \\ \midrule
\multirow{6}{*}[-31pt]{\textbf{\begin{tabular}[c]{@{}c@{}}Static\\ ($\mathcal{F}_{S}$)\end{tabular}}} & ID of task-j in the DAG & Task \\ \cmidrule(l){2-3} 
 & \begin{tabular}[c]{@{}l@{}}Execution time of a task $T_{j}$\\ in PE $P_{i}$  ($t_{exe}(P_i,T_j)$)\end{tabular} & \begin{tabular}[c]{@{}c@{}}Task\\ PE\end{tabular} \\ \cmidrule(l){2-3} 
 & \begin{tabular}[c]{@{}l@{}}Downward depth of task $T_{j}$ \\ in the DAG\end{tabular} & \begin{tabular}[c]{@{}c@{}}Task\\ Application\end{tabular} \\ \cmidrule(l){2-3} 
 & \begin{tabular}[c]{@{}l@{}}IDs of predecessor tasks\\ of task $T_{j}$\end{tabular} & \begin{tabular}[c]{@{}c@{}}Task\\ Application\end{tabular} \\ \cmidrule(l){2-3} 
 & Application ID & Application \\ \cmidrule(l){2-3} 
 & \begin{tabular}[c]{@{}l@{}}Power consumption of task $T_{j}$\\ in PE $P_{i}$\end{tabular} & \begin{tabular}[c]{@{}c@{}}Task\\ PE\end{tabular} \\ \midrule
\multirow{4}{*}[-25pt]{\textbf{\begin{tabular}[c]{@{}c@{}}Dynamic\\ ($\mathcal{F}_{D}$)\end{tabular}}} & \begin{tabular}[c]{@{}l@{}}Relative order of task $T_{j}$ in \\ the ready queue\end{tabular} & Task \\ \cmidrule(l){2-3} 
 & \begin{tabular}[c]{@{}l@{}}Earliest time when PEs \\ in a cluster-$c$ are ready \\ for task execution\end{tabular} & PE \\ \cmidrule(l){2-3} 
 & \begin{tabular}[c]{@{}l@{}}Clusters in which predecessor \\ tasks of task $T_{j}$ executed\end{tabular} & Task \\ \cmidrule(l){2-3} 
 & \begin{tabular}[c]{@{}l@{}}Communication volume from task \\ $T_{j}$ to task $T_{k} (v_{jk})$\end{tabular} & \begin{tabular}[c]{@{}c@{}}Task\end{tabular} \\ \bottomrule
\end{tabular}
\end{table}

In summary, the features of a task comprehensively represent the task itself and the state of the processing elements in the system to effectively learn the decisions from the Oracle policy.
The specific types of features used in this work to represent the state
and their categories are listed in Table~\ref{tab:feature_selection_2}.
The static and dynamic features are denoted as $\mathcal{F}_S$ 
and $\mathcal{F}_D$, respectively. Then, we define the systems state at a given time instant $k$ using the features in Table~\ref{tab:feature_selection_2} as:
\begin{equation} \label{eq:state}
    s_k = \mathcal{F}_{S,k} \cup \mathcal{F}_{D,k}
\end{equation}
where $\mathcal{F}_{S,k}$ and $\mathcal{F}_{D,k}$ denote the static and dynamic features respectively at a given time instant $k$.
For an SoC with 16 processing elements grouped as 5 clusters, we obtain a set of 45 features for the proposed IL technique.


\subsection{Oracle Generation}
\label{ssec:oracle}

The goal of this work is to develop generalized scheduling models for streaming applications of multiple types to be executed on heterogeneous many-core systems.
The generality of the IL-based scheduling framework enables using IL with any Oracle.
The Oracle can be any scheduling algorithm that optimizes an arbitrary metric, such as execution time, power consumption, and total SoC energy.

To generate the training dataset,
we implemented both optimal schedulers using CP and heuristics. 
These schedulers are integrated into a SoC simulation framework, as explained under experimental results. 
Suppose a new task $T_j$ becomes ready at time $k$. 
The Oracle is called to schedule the task to a PE. 
The Oracle policy for this action task with system state $s_k$ can be expressed as:
\begin{equation}
    \pi^*(s_k) = P_i, 
\end{equation}
where $P_i \in \mathcal{P}$ is the PE $T_j$ scheduled to 
and $s_k$ is the system state defined in Equation~\ref{eq:state}.
After each scheduling action, the particular task that is scheduled ($T_j$), 
the system state $s_k \in \mathcal{S}$, and the scheduling decision 
are added to the training data.
To enable the Oracle policies to generalize for different workload conditions, 
we constructed workload mixes using the target applications at different data rates, as detailed in Section~\ref{ssec:experimental_setup}. 

\subsection{IL-based Scheduling Framework}
\label{ssec:il_framework}

\begin{algorithm}[t]
\footnotesize
\SetNoFillComment
\caption{Hierarchical imitation learning Framework}
\label{algo:hierarchical}
\SetAlgoLined
\For {task T $\in \mathcal{T}$} {
$s$ = Get current state for task \textit{T} \\
\tcc{\textbf{Level-1 IL policy to assign cluster}}
$c$ = $\pi_{C}(s)$ \\
\tcc{\textbf{Level-2 IL policy to assign PE}}
p = $\pi_{P,c}(s)$ \\
\tcc{\textbf{Assign \textit{T} to the predicted PE}}
}
\end{algorithm}

This section presents the hierarchical IL-based scheduler for runtime task scheduling in heterogeneous many-core platforms.
A hierarchical structure is more scalable since it breaks a complex scheduling problem down into simpler problems.
Furthermore, it achieves a significantly higher classification accuracy compared to a flat classifier~(>93\% versus 55\%), 
as detailed in Section~\ref{ssec:il_performance}. 

Our hierarchical IL-based scheduler policies approximate the Oracle with two levels, as outlined in Algorithm~\ref{algo:hierarchical}.
The first level policy $\pi_{C}(s): \mathcal{S} \rightarrow \mathcal{C}$ is a coarse-grained scheduler 
that assigns tasks into clusters. 
This is a natural choice since individual PEs within a cluster have identical static parameters, 
i.e., they differ only in terms of their dynamic states. 
The second level (i.e., fine-grained scheduling)
consists of \textit{one dedicated policy} $\pi_{P,c}(s): \mathcal{S} \rightarrow \mathcal{P}$ 
for each cluster $c \in \mathcal{C}$. 
These policies assign the input task to a PE within its own cluster, 
i.e., $\pi_{P,c}(s) \in \mathcal{P}^c, ~\forall c \in \mathcal{C}$.
We leverage off-the-shelf machine learning techniques, such as regression trees and neural networks, to construct the IL policies.
The application of these policies approximates the corresponding Oracle policies constructed offline.

IL policies suffer from error propagation as the state-action pairs in the Oracle are not necessarily i.i.d. (independent and identically distributed).
Specifically, if the decision taken by the IL policies at a particular decision epoch is different from the Oracle, then the resultant state for the next epoch is also different with respect to the Oracle.
Therefore, the error further accumulates at each decision epoch.
This can occur during runtime task scheduling
when the policies are applied to applications that the policies did not train with.
This problem is addressed by the data aggregation algorithm (DAgger), 
proposed to improve IL policies~\cite{ross2011reduction}.
DAgger adds the system state and the Oracle decision to the training data whenever the IL policy makes a wrong decision.
Then, the policies are retrained after the execution of the workload.

\begin{algorithm}[tb]
\footnotesize
\SetNoFillComment
\caption{Methodology to aggregate data in a hierarchical imitation learning framework}
\label{algo:dagger}
\SetAlgoLined
\For {task T $\in \mathcal{T}$} {
$s$ = Get current state for task T \\
\If {$\pi_{C}(s)$ == $\pi_{C}^*(s)$} {
    \If {$\pi_{P,c}(s)$ != $\pi_{P,c}^*(s)$} {
        Aggregate state $s$ and label $\pi_{P,c}^*(s)$ to the dataset
    }
}
\Else {
    Aggregate state $s$ and label $\pi_{C}^*(s)$ to the dataset \\
    $c^{*}$ = $\pi_{C}^*(s)$ \\
    \If {$\pi_{P,c^*}(s)$ != $\pi_{P,c^{*}}^*(s)$} {
        Aggregate state $s$ and label $\pi_{P,c}^*(s)$ to the dataset
    }
}
\tcc{\textbf{Assign T to the predicted PE}}
}
\end{algorithm}

DAgger is not readily applicable to the runtime scheduling problem 
since the number of states is unbounded as a scheduling decision at time $t$ for state $s$ ($s_{t}$) can result in any possible resultant state, $s_{t+1}$.
In other words, the feature space is continuous, and hence, it is infeasible to generate an exhaustive Oracle offline.
We overcome this challenge by generating an Oracle on-the-fly.
More specifically, we incorporate the proposed framework into a simulator. 
The offline scheduler used as the Oracle is called \textit{dynamically} 
for each new task. 
Then, we augment the training data with all the features, Oracle actions, as well as the results of the IL policy under construction. 
Hence, the data aggregation process is performed as part of the dynamic simulation.

The hierarchical nature of the proposed IL framework introduces one more complexity to data aggregation. 
The cluster policy's output may be correct, while the PE cluster reaches a wrong decision (or vice versa).
If the cluster prediction is correct, we use this prediction to select the PE policy of that cluster, as outlined in Algorithm~\ref{algo:dagger}.
Then, if the PE prediction is also correct, the execution continues; otherwise, the PE data is aggregated in the dataset. 
However, if the cluster prediction does not align with the Oracle, in addition to aggregating the cluster data, an on-the-fly Oracle is invoked to select the PE policy, then the PE prediction is compared to the Oracle, and the PE data is aggregated in case of a wrong prediction.

\section{Experimental Results}
\label{sec:experimental_results}


\new{Section~\ref{ssec:experimental_setup} presents the experimental methodology and setup.
Section~\ref{ssec:ml_techniques} explores different machine learning classifiers for IL.
The significance of the proposed features is studied using a regression tree classifier in Section~\ref{ssec:feature_exploration}.
Section~\ref{ssec:il_performance} presents the evaluation of the proposed IL-scheduler.
Section~\ref{ssec:generalization} analyzes the generalization capabilities of IL-scheduler.
The performance analysis with multiple workloads is presented in Section~\ref{ssec:multiworkload}.
We demonstrate the evaluation of the proposed IL technique to energy-based optimization objectives in Section~\ref{ssec:energy}.
Section~\ref{ssec:rl} presents comparisons with RL-based scheduler and Section~\ref{ssec:complexity} analyzes the complexity of the proposed approach.
}

\subsection{Experimental Methodology and Setup}
\label{ssec:experimental_setup}

\renewcommand{\arraystretch}{0.95}
\begin{table}[b]
\centering
\caption{Characteristics of applications used in this study and the number of frames of each application in the workload}
\label{tab:apps}
\setlength\tabcolsep{3.5pt}
\begin{tabular}{@{}rcclcl@{}}
\toprule
\multicolumn{1}{c}{\multirow{2}{*}{\textbf{App}}} & \multirow{2}{*}{\textbf{\begin{tabular}[c]{@{}c@{}}\# of \\ Tasks\end{tabular}}} & \multirow{2}{*}{\textbf{\begin{tabular}[c]{@{}c@{}}Execution\\ Time ($\mu$s)\end{tabular}}} & \multicolumn{1}{c}{\multirow{2}{*}{\textbf{\begin{tabular}[c]{@{}c@{}}Supported \\ Clusters\end{tabular}}}} & \multicolumn{2}{c}{\textbf{\begin{tabular}[c]{@{}c@{}}Representation\\ in workload  \end{tabular}}} \\ \cmidrule(l){5-6} 
\multicolumn{1}{c}{} &  &  & \multicolumn{1}{c}{} & \textbf{\begin{tabular}[c]{@{}c@{}}\#frames\end{tabular}} & \multicolumn{1}{c}{\textbf{\begin{tabular}[c]{@{}c@{}}\#tasks\end{tabular}}} \\ \midrule
WiFi-TX & 27 & 301 & \begin{tabular}[c]{@{}l@{}}big, LITTLE, FFT\end{tabular} & 69 & 1863 \\ \midrule
WiFi-RX & 34 & 71 & \begin{tabular}[c]{@{}l@{}}big, LITTLE, \\ FFT, Viterbi\end{tabular} & 111 & 3774 \\ \midrule
RangeDet & 7 & 177 & \begin{tabular}[c]{@{}l@{}}big, LITTLE, FFT\end{tabular} & 64 & 448 \\ \midrule
SC-TX & 8 & 56 & big, LITTLE & 64 & 512 \\ \midrule
SC-RX & 8 & 154 & \begin{tabular}[c]{@{}l@{}}big, LITTLE, \\ Viterbi\end{tabular} & 91 & 728 \\ \midrule
TempMit & 10 & 81 & \begin{tabular}[c]{@{}l@{}}big, LITTLE, \\ Matrix mult.\end{tabular} & 101 & 1010 \\ \midrule
\multicolumn{4}{l}{\textbf{TOTAL}} & \textbf{500} & \multicolumn{1}{l}{\textbf{8335}} \\ \bottomrule
\end{tabular}
\end{table}

\noindent\textbf{\textit{Domain Applications}:}
The proposed IL scheduling methodology is evaluated using applications from wireless communication and radar processing domains.
We employ WiFi-transmitter (\textit{WiFi-TX}), WiFi-receiver (\textit{WiFi-RX}), range detection (\textit{RangeDet}), single-carrier transmitter (\textit{SC-TX}), single-carrier receiver (\textit{SC-RX}) and temporal mitigation (\textit{TempMit}) applications, 
as summarized in Table~\ref{tab:apps}.
We construct workload mixes using these applications and run them in parallel. 

\begin{figure}[t]
	\centering
	\includegraphics[width=0.6\linewidth]{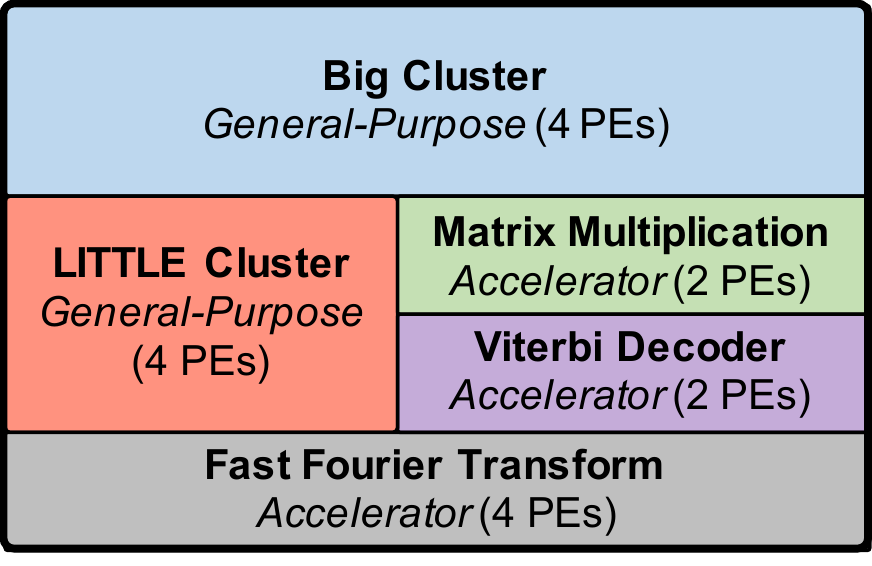}
	\caption{Configuration of the heterogeneous many-core platform comprising 16 processing elements, used for scheduler evaluations.}
	\label{fig:soc}
\end{figure}

\noindent\textbf{\textit{Heterogeneous DSSoC Configuration}:}
Considering the nature of applications, we employ a 
DSSoC with 16 PEs, including accelerators for the most computationally intensive tasks; they are divided into five clusters with multiple homogeneous PEs, as illustrated in Fig.~\ref{fig:soc}.
To enable power-performance trade-off while using general-purpose cores, 
we include a big cluster with four Arm A57 cores and a LITTLE cluster with four Arm A53 cores. 
In addition, the DSSoC integrates accelerator clusters for matrix multiplication, FFT, and Viterbi decoder to address the computing requirements of the target domain applications summarized in Table~\ref{tab:apps}. 
The accelerator interfaces are adapted from~\cite{mack2020user}.
The number of accelerator instances in each cluster is selected based on how much the target applications use them. 
For example, three out of the six reference applications involve FFT, while range detection application alone has three FFT operations.
Therefore, we employ four instances of FFT hardware accelerators and two instances of Viterbi and matrix multiplication accelerators, as shown in Fig.~\ref{fig:soc}.

\noindent{\textbf{\textit{Simulation Framework}:}}
We evaluate the proposed IL scheduler using the discrete event-based simulation framework~\cite{arda2020ds3}, \textit{which is validated against two commercial SoCs}: Odroid-XU3~\cite{ODROID_Platforms} and Zynq Ultrascale+ ZCU102~\cite{FPGA}.
This framework enables simulations of the target applications modeled as DAGs under different scheduling algorithms. 
More specifically, a new instance of a DAG arrives following a 
specified inter-arrival time rate and distribution, such as an exponential distribution. 
After the arrival of each DAG instance, called a frame, 
the simulator calls the scheduler under study. 
Then, the scheduler uses the information in the DAG and the current system state to assign the ready tasks to the waiting queues of the PEs. 
The simulator facilitates storing this information and the scheduling decision to construct the Oracle, as described in Section~\ref{ssec:oracle}. 

The execution times and power consumption for the tasks in our domain applications are profiled on Odroid-XU3 and Zynq ZCU102 SoCs. The simulator uses these profiling results to determine the execution time and power consumption of each task.
After all the tasks that belong to the same frame are executed, 
the processing of the corresponding frame completes.
The simulator keeps track of the execution time and energy consumed for each frame. 
These end-to-end values are within 3\%, on average, of the measurements on Odroid-XU3 and Zynq ZCU102 SoCs.

\begin{figure}[t]
	\centering
	\includegraphics[width=1.0\linewidth]{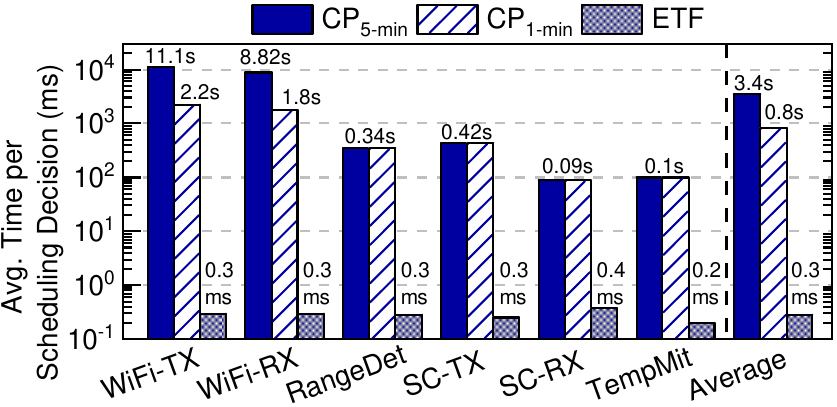}
	\caption{A comparison of average runtime per scheduling decision for each application with $CP_{5-min}$, $CP_{1-min}$ and ETF schedulers.}
	\label{fig:cp_runtimes}
\end{figure} 

\noindent{\textbf{\textit{Scheduling Algorithms used for Oracle and Comparisons}:}}
We developed a CP formulation using IBM ILOG CPLEX Optimization Studio~\cite{cplex} to obtain the optimal schedules whenever the problem size allows. 
After the arrival of each frame, the simulator calls the CP solver to find the schedule dynamically as a function of the current system state.
Since the CP solver takes hours for large inputs ($\sim$100 tasks), 
we implemented two versions with one minute (CP$_{1-min}$) and five minutes (CP$_{5-min}$) time-out per scheduling decision. When the model fails to find an optimal schedule, we use the best solution found within the time limit.
Fig.~\ref{fig:cp_runtimes} shows that the average time of the CP solver per scheduling decision 
for the benchmark applications is about 0.8 seconds and 3.5 seconds, respectively, based on the time limit. 
Consequently, one entire simulation can take up to 2 days, even with a time-out.

We also implemented the ETF heuristic scheduler, which goes over all tasks and possible assignments to find the earliest finish time considering communication overheads. 
Its average execution time is close to 0.3~ms, which is still prohibitive for a runtime scheduler, as shown in Fig.~\ref{fig:cp_runtimes}.
However, we observed that it performs better than CP$_{1-min}$ and marginally worse than 
CP$_{5-min}$, as we detail in Section~\ref{ssec:il_performance}.

Oracle generation with the CP formulation is not practical for two reasons. First, it is possible that for small input sizes (e.g., less than ten tasks), there might be multiple (incumbent) optimal solutions, and CP would choose one of them randomly.
The other reason is that for large input sizes, CP terminates at the time limit providing the best solution found so far, which is sub-optimal.
The sub-optimal solutions produced by CP vary based on the problem size and the limit.
In contrast, ETF is easier to imitate at runtime 
and its results are within 8.2\% of CP$_{5-min}$ results. 
Therefore, we use ETF as the Oracle policy in our experiments and use the results of CP schedulers as reference points. 
We train IL policies for this Oracle in Section~\ref{ssec:ml_techniques} and evaluate their performance in Section~\ref{ssec:il_performance}.



\begin{table}[t]
\centering
\caption{Classification accuracies of trained IL policies with different machine learning classifiers.}
\label{tab:ml_accuracies}
\setlength\tabcolsep{3.5pt}
\begin{tabular}{@{}cc|ccccc@{}}
\toprule
\textbf{Classifier} & \textbf{\begin{tabular}[c]{@{}c@{}}Cluster\\ Policy\end{tabular}} & \textbf{\begin{tabular}[c]{@{}c@{}}LITTLE\\ Policy\end{tabular}} & \textbf{\begin{tabular}[c]{@{}c@{}}big \\ Policy\end{tabular}} & \textbf{\begin{tabular}[c]{@{}c@{}}MatMult\\ Policy\end{tabular}} & \textbf{\begin{tabular}[c]{@{}c@{}}FFT\\ Policy\end{tabular}} & \textbf{\begin{tabular}[c]{@{}c@{}}Viterbi \\ Policy\end{tabular}} \\ \midrule
\textbf{RT} & 99.6 & \textbf{93.8} & \textbf{95.1} & 99.9 & 99.5 & 100 \\
\textbf{SVC} & 95.0 & \textbf{85.4} & \textbf{89.9} & 97.8 & 97.5 & 98.0 \\
\textbf{LR} & 89.9 & \textbf{79.1} & \textbf{72.0} & 98.7 & 98.2 & 98.0 \\
\textbf{NN} & 97.7 & \textbf{93.3} & \textbf{93.6} & 99.3 & 98.9 & 98.1 \\ \bottomrule
\end{tabular}
\end{table}

\begin{table}[t]
\centering
\caption{Execution time and storage overheads per IL policy for regression tree and neural network classifiers.}
\label{tab:ml_overheads}
\setlength\tabcolsep{3.5pt}
\begin{tabular}{@{}cccc@{}}
\toprule
\multirow{2}{*}{\textbf{Classifier}} & \multicolumn{2}{c}{\textbf{Latency ($\mu$s)}} & \multirow{2}{*}{\textbf{Storage (KB)}} \\ \cmidrule(lr){2-3}
 & \textbf{\begin{tabular}[c]{@{}c@{}}Odroid-XU3\\ (Arm A15)\end{tabular}} & \textbf{\begin{tabular}[c]{@{}c@{}}Zynq Ultrascale+ ZCU102\\ (Arm A53)\end{tabular}} &  \\ \midrule
\textbf{RT} & 1.1 & 1.1 & 19.3 \\
\textbf{NN} & 14.4 & 37 & 16.9 \\ \bottomrule
\end{tabular}
\end{table}

\subsection{Exploring Different Machine Learning Classifiers for IL}
\label{ssec:ml_techniques}
We explore various ML classifiers within the IL methodology to approximate the Oracle policy.
One of the key metrics that drive the choice of machine learning techniques is the classification accuracy of the \new{IL} policies.
At the same time, the policy should also have a low storage and execution time overheads.
We evaluate the following algorithms for classification accuracy and implementation efficiency: regression tree (RT), support vector classifier (SVC), logistic regression (LR), and a multi-layer perceptron neural network (NN) with 4 hidden layers and 32 neurons in each hidden layer.

The classification accuracy of ML algorithms under study are listed in Table~\ref{tab:ml_accuracies}.
In general, all classifiers achieve a high accuracy to choose the cluster (the first column). At the second level, they choose the correct PE with high accuracy (>97\%) within the hardware accelerator clusters. 
However, they have lower accuracy and larger variation for the LITTLE and big clusters (highlighted columns).
This is intuitive as the LITTLE and big clusters can execute all types of tasks in the applications, whereas accelerators execute fewer tasks.
In strong contrast, a flat policy, which directly predicts the PE,
results in training accuracy with 55\% at best.
Therefore, we focus on the proposed hierarchical IL methodology.

\begin{table}[t]
\centering
\setlength\tabcolsep{2pt}
\setlength\extrarowheight{0.6pt}
\captionof{table}{\new{Training accuracy of IL policies with subsets of the proposed feature set}}
\begin{tabular}{@{}lcccccc@{}}
\toprule
\multicolumn{1}{c}{\textbf{\begin{tabular}[c]{@{}c@{}}Features Excluded \\ from Training\end{tabular}}} & \textbf{\begin{tabular}[c]{@{}c@{}}Cluster \\ Policy\end{tabular}} & \textbf{\begin{tabular}[c]{@{}c@{}}LITTLE \\ Policy\end{tabular}} & \textbf{\begin{tabular}[c]{@{}c@{}}big\\ Policy\end{tabular}} & \textbf{\begin{tabular}[c]{@{}c@{}}MatMul\\ Policy\end{tabular}} & \textbf{\begin{tabular}[c]{@{}c@{}}FFT\\ Policy\end{tabular}} & \textbf{\begin{tabular}[c]{@{}c@{}}Viterbi\\ Policy\end{tabular}} \\ \midrule
\textbf{None} & \textbf{99.6} & \textbf{93.8} & \textbf{95.1} & \textbf{99.9} & \textbf{99.5} & \textbf{100} \\
Static features & 87.3 & 93.8 & 92.7 & 99.9 & 99.5 & 100 \\
Dynamic features & 88.7 & 52.1 & 57.6 & 94.2 & 70.5 & 98 \\
PE availability times & 92.2 & 51.1 & 61.5 & 94.1 & 66.7 & 98.1 \\
Task ID, depth, app. ID & 90.9 & 93.6 & 95.3 & 99.9 & 99.5 & 100 \\ \bottomrule
\end{tabular}
\label{tab:feature_exploration}
\end{table}

Regression trees (RT) trained with a maximum depth of 12 produce the best accuracy for the cluster and PE policies, with more than 99.5\% accuracy for the cluster and hardware acceleration policies.
RT also produces an accuracy of 93.8\% and 95.1\% to predict PEs within the LITTLE and big clusters, respectively, which is the highest among all the evaluated classifiers.
The classification accuracy of NN policies are comparable to RT, 
with a slightly lower cluster prediction accuracy of 97.7\%.
In contrast, support vector classifiers (SVC) and logistic regression (LR) are not preferred due to lower accuracy of less than 90\% and 80\%, respectively, to predict PEs within LITTLE and big clusters.

\begin{figure}[b]
    \centering  
    \includegraphics[width=1.0\linewidth]{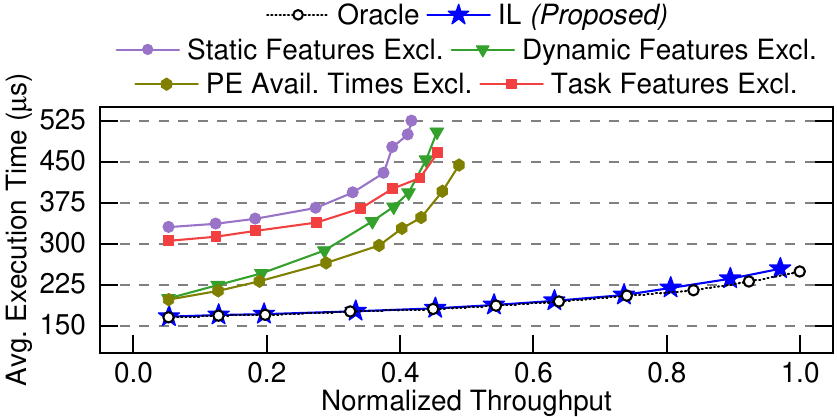}
    \captionof{figure}{\new{Average execution time comparison of the applications with Oracle, IL (\textit{Proposed}) and IL policies with subsets of features. As shown, the average execution time with IL closely follows the Oracle.}}
    \label{fig:feature_exploration}
\end{figure} 

We choose regression trees and neural networks to analyze the latency and storage overheads (due to their superior performance).
The latency of RT is 1.1$\mu$s on Arm Cortex-A15 in Odroid-XU3 and on Arm Cortex-A53 in Zynq ZCU102, as shown Table~\ref{tab:ml_overheads}. 
In comparison, the scheduling overhead of CFS, the default Linux scheduler, on Zynq ZCU102 running Linux Kernel 4.9 is 1.2$\mu$s, which is slightly larger than our solution. 
The storage overhead of an RT policy is 19.33~KB.
The NN policies incur an overhead of 14.4$\mu$s on the Arm Cortex-A15 cluster in Odroid-XU3 and 37$\mu$s on Arm Cortex-A53 in Zynq, with a storage overhead of 16.89 KB.
NNs are preferable for use in an online environment as their weights can be incrementally updated using the back-propagation algorithm.
However, due to competitive classification accuracy and lower latency overheads of RTs over NNs, we choose RT for the rest of the experiments.

\subsection{\new{Feature Space Exploration with Regression Tree Classifier}}
\label{ssec:feature_exploration}

\new{This section explores the significance of the features chosen to represent the state. 
For this analysis, 
we assess the impact of the input features on the training accuracy with RT classifier and average execution time following a systematic approach.}

\new{The training accuracy with subsets of features and the corresponding scheduler performance is shown in Table~\ref{tab:feature_exploration} and Fig.~\ref{fig:feature_exploration} respectively.
\textit{First}, we exclude all static features from the training dataset. The training accuracy for the prediction of the cluster significantly drops by 10\%. 
Since we use hierarchical IL policies, an incorrect first-level decision results in a significant penalty for the decisions at the next level.
\textit{Second}, we exclude all dynamic features from training. 
This results in a similar impact for the cluster policy (10\%) but significantly affects the policies constructed for the LITTLE, big, and FFT.
\textit{Next}, a similar trend is observed when PE availability times are excluded from the feature set.
The accuracy is marginally higher since the other dynamic features contribute to learning the scheduling decisions.
\textit{Finally}, we remove a few task related features such as the downward depth, task, and application identifier. In this case, the impact is to the cluster policy accuracy since these features describe the node in the DAG and influence the cluster mapping.
}

\new{As observed in Fig.~\ref{fig:feature_exploration}, the average execution time of the workload significantly degrades when all features are not included.
Hence, the chosen features help to construct effective IL policies, approximating the Oracle with over 99\% accuracy in execution time.}

\begin{figure}[b]
	\centering
	\includegraphics[width=1.0\linewidth]{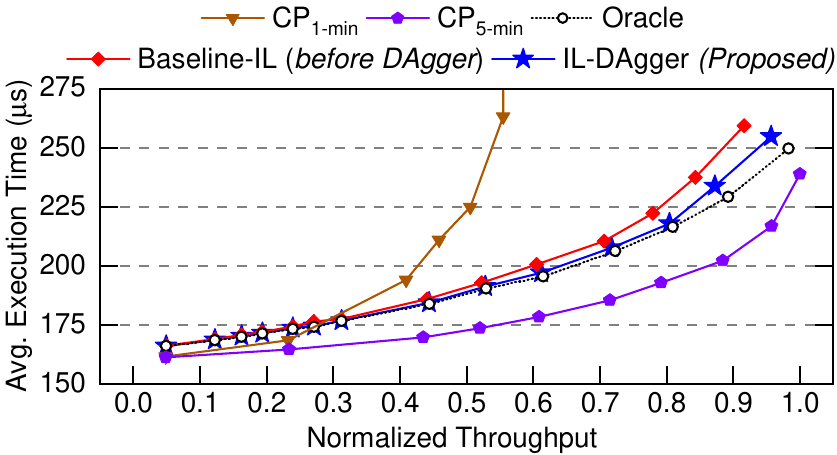}
	\caption{Comparison of average job execution time between Oracle, CP solutions, and imitation learning policies to schedule a workload comprising a mix of six streaming applications. IL scheduler policies with baseline-IL (\textit{before DAgger}) and with IL-DAgger (\textit{Proposed}) are shown in the comparison.}
	\label{fig:basic_dagger}
\end{figure}

\subsection{IL-Scheduler Performance Evaluation}
\label{ssec:il_performance}

This section compares the performance of the proposed policy to the ETF Oracle, CP$_{1-min}$, and CP$_{5-min}$.
Since heterogeneous many-core systems are capable of running multiple applications simultaneously, 
we stream the frames in our application mix (see Table~\ref{tab:apps}) with increasing injection rates.
For example, a normalized throughput of 1.0 in Fig.~\ref{fig:basic_dagger} corresponds to 19.78 frames/ms. 
Since the frames are injected faster than they can be processed,
there are many overlapping frames at any given time. 

First, we train the IL policies with all six reference applications and refer to this as the baseline-IL scheduler.
IL policies suffer from error propagation due to the non i.i.d. nature of training data.
To overcome this limitation, we use a data aggregation technique adapted for a hierarchical IL framework (IL-DAgger), as discussed in Section~\ref{ssec:il_framework}.
A DAgger iteration involves executing the entire workload. 
We execute ten DAgger iterations and choose the best iteration with performance within 2\% of the Oracle.
If we fail to achieve the target, we continue to perform more iterations.

Fig.~\ref{fig:basic_dagger} shows that the proposed IL-DAgger scheduler 
performs almost identical to the Oracle; the mean average percentage difference between them is 1\%. 
More notably, the gap between the proposed IL-DAgger policy and 
the optimal CP$_{5-min}$ solution is only 9.22\%. 
We emphasize that CP$_{5-min}$ is included only as a reference point, 
but it has six orders of magnitude larger execution time overhead and cannot be used at runtime. 
Furthermore, the proposed approach performs better than CP$_{1-min}$, 
which is not able to find a good schedule within the one-minute time limit per decision. 
Finally, we note that the baseline IL can approach the performance of the proposed policy. 
This is intuitive since both policies are tested on known applications in this experiment. This is in contrast to the leave one out experiments presented in Section~\ref{ssec:generalization}. 

\noindent\new{{\textbf{\new{Pulse Doppler Application Case Study:}}
We demonstrate the applicability of the proposed IL-scheduling technique in complex scenarios using a pulse Doppler application. 
It is a real-world radar application, which computes the velocity of a moving target object. 
This application is significantly more complex, with 13-64$\times$ more tasks than the other applications. Specifically, it consists of 449 tasks comprising 192 FFT tasks, 128 inverse-FFT tasks, and 129 other computations.
The FFT and inverse-FFT operations can execute on the general-purpose cores and hardware accelerators. 
In contrast, the other tasks can execute only on the general-purpose cores. 
}}

\begin{figure}[b]
	\centering
	\includegraphics[width=1.0\linewidth]{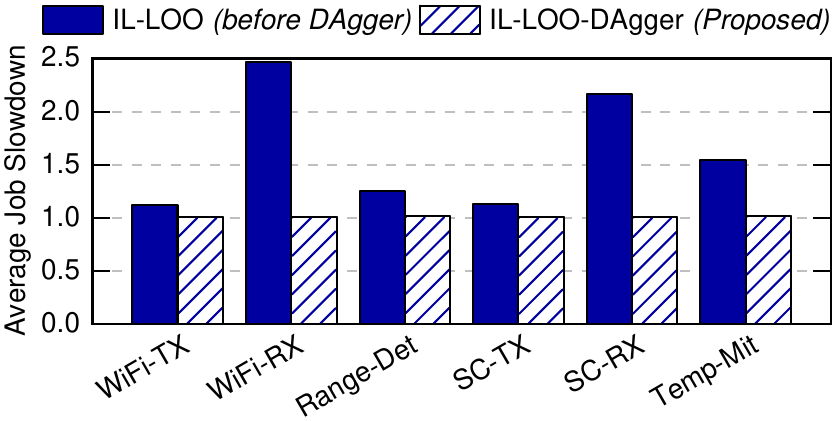}
	\caption{Average slowdown of IL policies in comparison with Oracle for leave-one-out (LOO) experiments before and after DAgger (\textit{Proposed}).}
	\label{fig:loo_slowdown}
\end{figure}

\begin{figure*}[!t]
	\centering
	\includegraphics[width=1.0\linewidth]{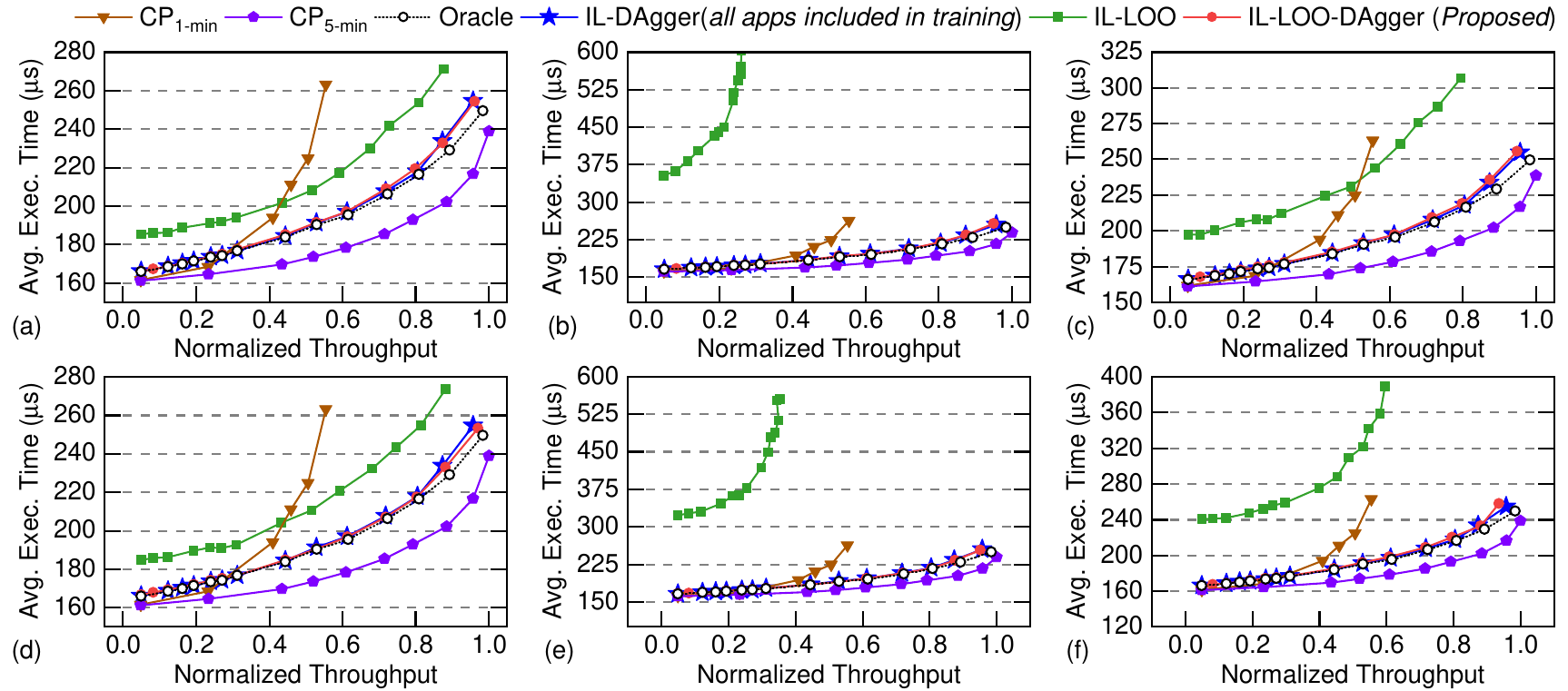}
	\captionof{figure}{Average execution time with Oracle, IL-DAgger (\textit{all applications are included for training}), IL with one application excluded from training (IL-LOO) and finally, the leave-one-out policy improved with DAgger (\textit{Proposed} IL-LOO-DAgger) technique. The \textit{excluded} applications are: (a) WiFi-TX, (b) WiFi-RX, (c) Range Detection (d) Single-Carrier TX (e) Single-Carrier RX and (f) Temporal Mitigation.}
	\label{fig:loo_dagger}
\end{figure*}

\new{The proposed IL policies achieve an average execution time within 2\% of the Oracle. The 2\% error is acceptable, considering that the application saturates the computing platform quickly due to its high complexity. Moreover, the CP-based approach does not produce a viable solution either with 1-minute or 5-minute time limits due to the large problem size. For this reason, this application is not included in our workload mixes and the rest of the comparisons. 
}

\subsection{\new{Illustration of Generalization with IL for Unseen Applications, Runtime Variations and Platforms}}
\label{ssec:generalization}
\new{This section analyzes the generalization of the proposed IL-based scheduling approach to unseen applications, runtime variations, and many-core platform configurations.}

\noindent \textbf{\new{IL-Scheduler Generalization to Unseen Applications using Leave-one-out Experiments:}}
IL, being an adaptation of supervised learning for sequential decision making, suffers from lack of generalization to unseen applications.
To analyze the effects of unseen applications, we train IL policies, excluding applications one each at a time from the training dataset~\cite{vehtari2017practical}.

To compare the performances of two schedulers $S_{1}$ and $S_{2}$, we use the job slowdown metric $slowdown_{S_1,S_2} = T_{S_{1}} / T_{S_{2}}$. ${Slowdown_{S_1,S_2}} > 1$ when $T_{S_{1}}>T_{S_{2}}$~\cite{mao2016resource}. 
The average slowdown of scheduler $S_{1}$ with respect to scheduler $S_{2}$ is computed as the average slowdown for all jobs at all injection rates.
The results present an interesting and intuitive explanation of the average job slowdown in execution times for each of the leave-one-out experiments.

Fig.~\ref{fig:loo_slowdown} shows the average slowdown of the baseline IL (IL-LOO) and proposed policy with DAgger iterations (IL-LOO-DAgger) with respect to the Oracle. 
We observe that the proposed policy outperforms the baseline IL for all applications, with the most significant gains obtained for WiFi-RX and SC-RX applications.
These two applications consist of a Viterbi decoder operation, which is very expensive to compute on general-purpose cores and highly efficient to compute on hardware accelerators.
When these applications are excluded, the IL policies are not exposed to the corresponding states in the training dataset and make incorrect decisions.
The erroneous PE assignments lead to an average slowdown of more than 2$\times$ for the receiver applications.
The slowdown when the transmitter applications (WiFi-TX and SC-TX) are excluded from training is approximately 1.13$\times$.
Range detection and temporal mitigation applications experience a slowdown of 1.25$\times$ and 1.54$\times$, respectively, for leave-one-out experiments.
The extent of the slowdown in each scenario depends on the application excluded from training and its execution time profile in the different processing clusters.
In summary, the average slowdown of all leave-one-out IL policies after DAgger (IL-LOO-DAgger) improves to \textasciitilde1.01$\times$ in comparison with the Oracle, as shown in Fig.~\ref{fig:loo_slowdown}.

Fig.~\ref{fig:loo_dagger}(a)-(f) show the average job execution times for the Oracle (ETF), baseline-IL, IL with leave-one-out and DAgger for IL with leave-one-out policies for each of the applications.
The highest number of DAgger iterations needed was 8 for SC-RX application, and the lowest was 2 for range detection application.
If the DAgger criterion is relaxed to achieving a slowdown of 1.02$\times$, all applications achieve the same in less than 5 iterations.
A drastic improvement in the accuracy of the IL policies with few iterations shows that the policies generalize quickly and well to unseen applications, thus making them suitable for applicability at runtime.

\noindent \textbf{\new{IL-Scheduler Generalization with Runtime Variations:}}
\begin{table}[t]
\centering
\setlength\tabcolsep{2.2pt}
\caption{\new{Standard deviation (in percentage of execution time) profiling of applications in Odroid-XU3 and Zynq ZCU-102.}}
\begin{tabular}{@{}ccccccc@{}}
\toprule
\textbf{Application} & \textbf{WiFi-TX} & \textbf{WiFi-RX} & \textbf{RangeDet} & \textbf{SC-TX} & \textbf{SC-RX} & \textbf{TempMit} \\ \midrule
Zynq ZCU-102 & 0.34 & 0.56 & 0.66 & 1.15 & 1.80 & 0.63 \\
Odroid-XU3 & 6.43 & 5.04 & 5.43 & 6.76 & 7.14 & 3.14 \\ \bottomrule
\end{tabular}
\label{tab:std_deviation}
\end{table}

\begin{table}[b]
\centering
\caption{\new{Configuration of many-core platforms.}}
\label{tab:soc_exploration}
\begin{tabular}{@{}cccccc@{}}
\toprule
\textbf{\begin{tabular}[c]{@{}c@{}}Platform \\ Config.\end{tabular}} & \textbf{\begin{tabular}[c]{@{}c@{}}LITTLE \\ PEs\end{tabular}} & \textbf{\begin{tabular}[c]{@{}c@{}}big\\ PEs\end{tabular}} & \textbf{\begin{tabular}[c]{@{}c@{}}MatMul\\ Acc. PEs\end{tabular}} & \textbf{\begin{tabular}[c]{@{}c@{}}FFT\\ Acc. PEs\end{tabular}} & \textbf{\begin{tabular}[c]{@{}c@{}}Decoder\\ Acc. PEs\end{tabular}} \\ \midrule
G1 (Baseline) & 4 & 4 & 2 & 4 & 2 \\
G2 & 2 & 2 & 2 & 2 & 2 \\
G3 & 1 & 1 & 1 & 1 & 1 \\
G4 & 4 & 4 & 1 & 1 & 1 \\
G5 & 4 & 4 & 0 & 0 & 0 \\ \bottomrule
\end{tabular}
\end{table}

\new{Tasks experience runtime variations due to variations in system workload, memory, and congestion. Hence, it is crucial to analyze the performance of the proposed approach when tasks experience such variations, rather than observing only their static profiles. 
Our simulator accounts for variations by using a Gaussian distribution to generate variations in execution time~\cite{xian2008dynamic}. 
To allow evaluation in a realistic scenario, all tasks in every application are profiled on big and LITTLE cores of Odroid-XU3, and, on Cortex-A53 cores and hardware accelerators on Zynq for variations in execution time.
}

\begin{figure}[t]
	\centering
	\includegraphics[width=1.0\linewidth]{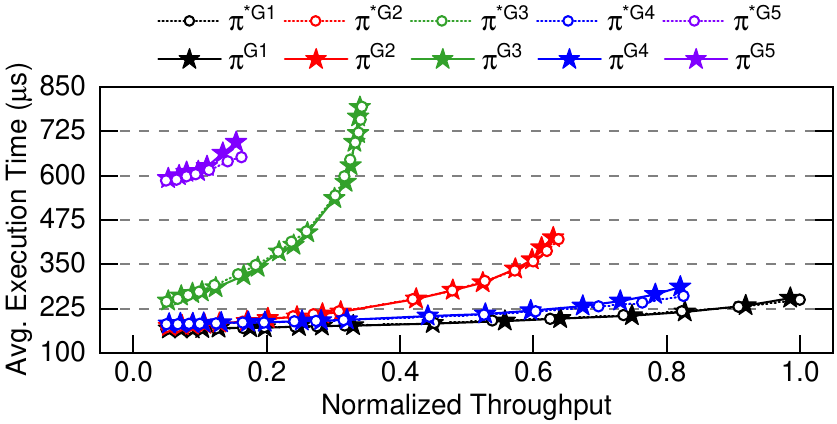}
	\caption{\new{IL policy evaluation with multiple many-core platform configurations. IL policies are trained with only configuration G1.}
    }
	\label{fig:soc_exploration}
\end{figure} 

\begin{figure*}[!t]
	\centering
	\includegraphics[width=1.0\linewidth]{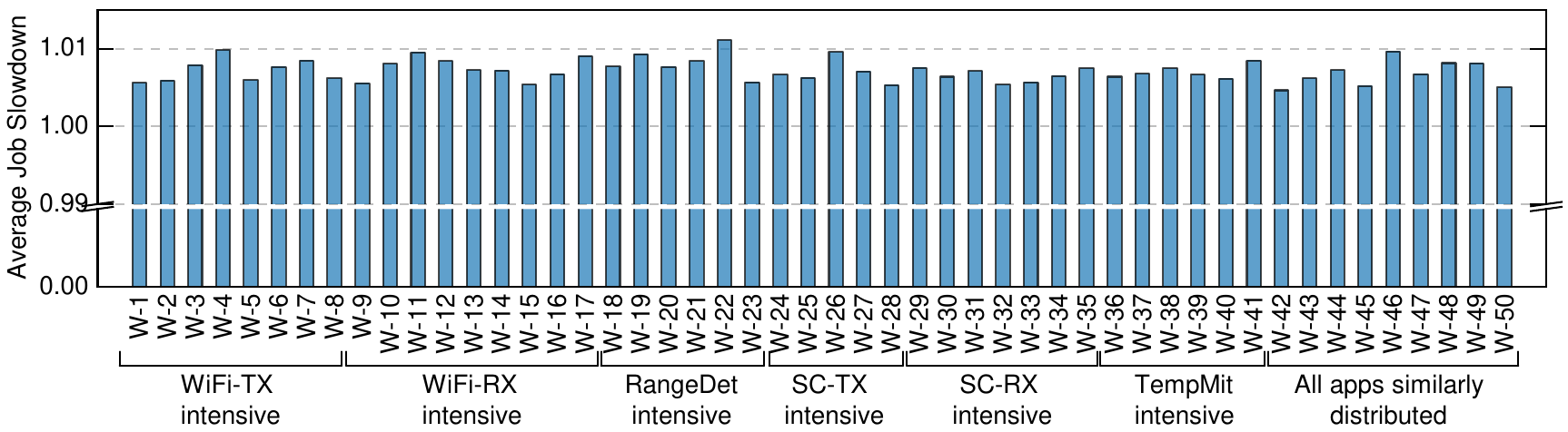}
	\caption{Comparison of average job slowdown normalized with IL-DAgger (\textit{Proposed}) policies against the Oracle for \textit{50} different workloads. The slowdown of IL-DAgger policies are shown for workloads with different \textit{intensities} of each application in the benchmark suite.}
	\label{fig:multiworkload}
\end{figure*} 

\new{We present the average standard deviation as a ratio of execution time for the tasks in Table~\ref{tab:std_deviation}.
The maximum standard deviation is less than 2\% of the execution time for the Zynq platform, and less than 8\% on the Odroid-XU3.
To account for variations in runtime, we add a noise of 1\%, 5\%, 10\%, and 15\% in task execution time during simulation.
The IL policies achieve average slowdowns of less than 1.01$\times$ in all cases of runtime variations.
Although IL policies are trained with static execution time profiles, the aforementioned results demonstrate that the IL policies adapt well to execution time variations at runtime.
Similarly, the policies also generalize to variations in communication time and power consumption.
}

\noindent \textbf{\new{IL-Scheduler Generalization with Platform Configuration:}} 
\new{This section presents a detailed analysis of the IL policies by varying the configuration i.e., the number of clusters, general-purpose cores, and hardware accelerators. 
To this end, we choose five different SoC configurations presented in Table~\ref{tab:soc_exploration}. 
The Oracle policy for a configuration G1 is denoted by $\pi^{*G1}$. 
An IL policy evaluated on configuration G1 is denoted as $\pi^{G1}$. 
}
\new{G1 is the baseline configuration that is used for extensive evaluation. 
Between configurations G1--G4, we vary the number of PEs within each cluster. 
We also consider a degenerate case that comprises only LITTLE and big clusters (configuration G5). 
We train IL policies with only configuration G1. 
The average execution times of $\pi^{G1}$, $\pi^{G2}$, and $\pi^{G3}$ are within 1\%, $\pi^{G4}$ performs within 2\%, and $\pi^{G5}$ performs within 3\%, of their respective Oracles. 
The accuracy of $\pi^{G5}$ with respect to the corresponding Oracle ($\pi^{*G5}$) is slightly lower (97\%) as the platform saturates the computing resources very quickly, as shown in Fig.~\ref{fig:soc_exploration}.
}


\begin{figure*}[!t]
	\centering
	\includegraphics[width=1.0\linewidth]{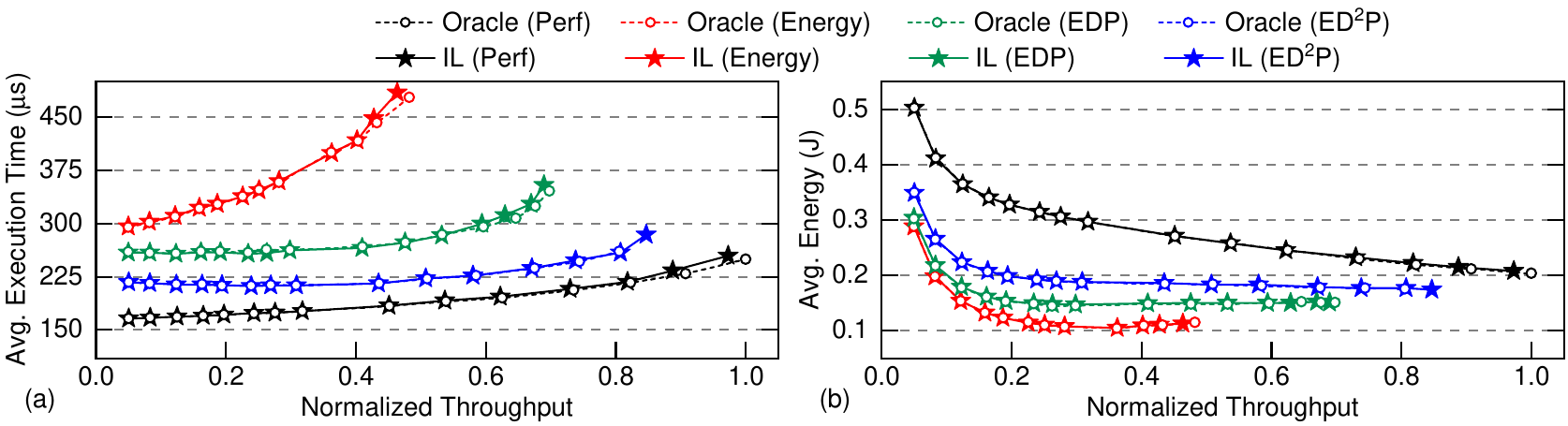}
	\captionof{figure}{\new{(a) Average execution time and (b) average energy consumption of the workload with Oracles and IL policies for performance, energy, energy-delay product (EDP) and energy-delay$^{2}$ product (ED$^{2}$P) objectives.}}
	\label{fig:graph_energy}
\end{figure*} 

\new{Based on these experiments, we observe that the IL policies generalize well for the different many-core platform configurations. 
The change in system configuration is accurately captured in the features (in execution times, PE availability times, etc.), which enables us to generalize well to new platform configurations.
When the cluster configuration in the many-core platform changes, the IL policies generalize well (within 3\%) but can also be improved by using DAgger to obtain improved performance (within 1\% of the Oracle).
}

\subsection{Performance Analysis with Multiple Workloads}
\label{ssec:multiworkload}

To demonstrate the generalization capability of the IL policies trained and aggregated on one workload (\textit{IL-DAgger}), we evaluate the performance of the same policies on \textit{50} different workloads consisting of different combinations of application mixes at varying injection rates, and each of these workloads contains 500 frames.
For this extensive evaluation, we consider workloads each of which are intensive on one of WiFi-TX, WiFi-RX, range detection, SC-TX, SC-RX, and temporal mitigation.
Finally, we also consider workloads in which all applications are distributed similarly.

Fig.~\ref{fig:multiworkload} presents the average slowdown for each of the \textit{50} different workloads (represented as \textit{W-1}, \textit{W-2} and so on).
While W-22 observes a slowdown of 1.01$\times$ against the Oracle, all other workloads experience an average slowdown of less than 1.01$\times$ (within 1\% of Oracle).
Independent of the distribution of the 
applications in the workloads, the IL policies approximate the Oracle well.
On average, the slowdown is less than 1.01$\times$, demonstrating the 
IL policies generalize
to different workloads and streaming intensities.

\subsection{\new{Evaluation with Energy and Energy-Delay Objectives}}
\label{ssec:energy}

\new{
Average execution time is crucial in configuring computing systems for meeting application latency requirements and user experience. 
Another critical metric in modern computing systems, especially battery-powered platforms, is energy consumption~\cite{moazzemi2019hessle,reddy2017inter}. 
Hence, this section presents the proposed IL-based approach with the following objectives: performance, energy, energy-delay product (EDP), and energy-delay$^{2}$ product (ED$^{2}$P). 
We adapt ETF to generate Oracles for each objective.
Then, the different Oracles are used to train IL policies for the corresponding objectives. 
The scheduling decisions are significantly more complex for these
Oracles. 
Hence, we use an RT of depth 16 (execution time uses RT of depth 12) to learn the decisions accurately. 
The average latency per scheduling decision remains similar for RT of depth 16 (\textasciitilde1.1$\mu$s) on Cortex-A53.
}

\new{Fig.~\ref{fig:graph_energy}(a) and Fig.~\ref{fig:graph_energy}(b) present the average execution time and average energy consumption, respectively, for IL policies with different objectives. 
The lowest energy is achieved by the energy Oracle, while it increases as more emphasis is added to performance (EDP $\rightarrow$ ED$^{2}$P $\rightarrow$ performance), as expected. 
The average execution time and energy consumption in all cases are within 1\% of the corresponding Oracles. This demonstrates the proposed IL scheduling approach is powerful as it learns from Oracles that optimize for any objective. 
}

\subsection{\new{Comparison with Reinforcement Learning}}
\label{ssec:rl}

\begin{figure}[t]
	\centering
	\includegraphics[width=1.0\linewidth]{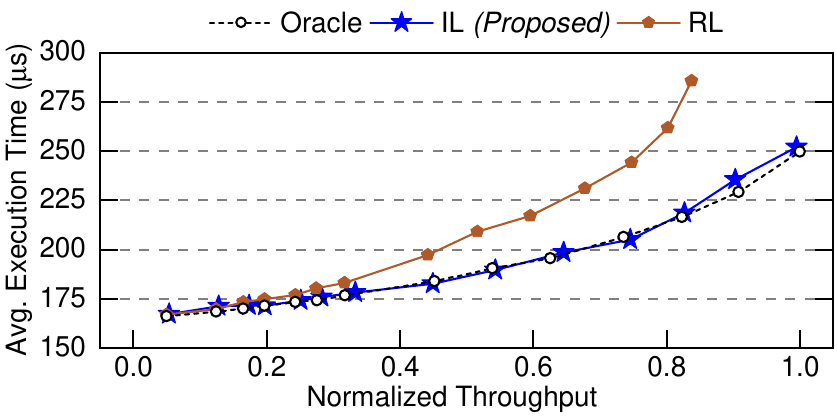}
	\caption{\new{Comparison of average execution time between Oracle, IL, and RL policies to schedule a workload comprising a mix of six streaming real-world applications.}
    }
	\label{fig:graph_rl}
\end{figure} 

Since the state-of-the-art machine learning techniques~\cite{mao2016resource,mao2019learning} do not target streaming DAG scheduling in heterogeneous many-core platforms, we implemented a policy-gradient based reinforcement learning technique using a deep neural network (multi-layer perceptron with 4 hidden layers with 32 neurons in each hidden layer) to compare with the proposed IL-based task scheduling technique. 
For the RL implementation, we vary the exploration rate between 0.01 to 0.99 and learning rate from 0.001 to 0.01. The reward function is adapted from~\cite{mao2019learning}. RL starts with random weights and then updates them based on the extent of exploration, exploitation, learning rate, and reward function. These factors affect convergence and quality of the learned RL models.

Fewer than 20\% of the experiments with RL  converge to a stable policy and less than 10\% of them provide competitive performance compared to the proposed IL-scheduler. 
We choose the RL solution that performs best to compare with the IL-scheduler.
The Oracle generation and training parts of the proposed technique take 5.6 minutes and 4.5 minutes, respectively, when running on an Intel Xeon E5-2680 processor at 2.40 GHz. In contrast, an RL-based scheduling policy that uses the policy gradient method converges in 300 minutes on the same machine.
Hence, the proposed technique is 30$\times$ faster than RL. 
As shown in Fig.~\ref{fig:graph_rl}, the RL scheduler performs within 11\% of the Oracle, whereas the IL scheduler presents average execution time that is within 1\% of the Oracle.

In general, RL-based schedulers suffer from the following drawbacks: (1) need for excessive fine-tuning of the parameters (learning rate, exploration rate, and NN structure), (2) reward function design, and (3) slow convergence for complex problems. 
In strong contrast, IL policies are guided by strong supervision eliminating the slow convergence problem and the need for a reward function.

\subsection{Complexity Analysis of the Proposed Approach}
\label{ssec:complexity}
In this section, we compare the complexity of our proposed IL-based task scheduling approach with ETF, which is used to construct the Oracle policies.
The complexity of ETF is $O(n^{2}m)$~\cite{hwang1989scheduling}, where $n$ is the number of tasks and $m$ is the number of PEs in the system.
While ETF is suitable for use in Oracle generation (offline), it is not efficient for online use due to the quadratic complexity on the number of tasks.
However, the proposed IL-policy which uses regression tree has the complexity of $O(n)$.
Since the complexity of the proposed IL-based policies is linear, it is practical to implement in heterogeneous many-core systems.

\section{Conclusion and Future Work} 
\label{sec:conclusion}
Efficient task scheduling in heterogeneous many-core platforms is crucial to improve the system performance, but is very challenging due to its NP-hardness.
In this work, we have presented an imitation learning based approach for task scheduling in many-core platforms executing streaming applications from wireless communications and radar systems.
We have presented a hierarchical imitation learning framework that learns from an Oracle to develop task scheduling policies to minimize the execution time of applications.
The framework has been evaluated comprehensively with six domain-specific applications and analyzed the storage and latency overheads of the IL policies.
We have shown that the IL policies approximate the Oracle better than 99\%.
The overhead of the policies are significantly low at 1.1$\mu$s latency per scheduling decision and lower than the completely fair scheduler (1.2$\mu$s).
Our IL policies achieve application execution times within 9.3\% of optimal schedules obtained offline using constraint programming.

Preliminary experiments in which we have used IL to bootstrap RL for task scheduling in heterogeneous many-core platforms, show much faster convergence to optimal policies.
We envision this work to initiate a new direction in scheduling studies with optimal Oracle generation and evaluation with applications from various domains.

\bibliographystyle{IEEEtranS}
{\footnotesize{\bibliography{references/scheduler}}}
\vskip -1.0 \baselineskip plus -1fil
\begin{IEEEbiography}
[{\includegraphics[width=1in,height=1.25in,clip,keepaspectratio]{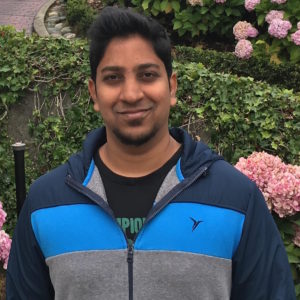}}]{Anish Krishnakumar} received his B.Tech degree in Electrical and Electronics Engineering from National Institute of Technology, Tiruchirappalli, India and Masters degree from Birla Institute of Technology, Pilani, India. Anish worked as a Physical Design Engineer at Qualcomm and as a Research Scientist at Intel Labs in India. He is currently pursuing Ph.D in Electrical Engineering at Arizona State University.
\end{IEEEbiography}
\vskip -1.0 \baselineskip plus -1fil
\begin{IEEEbiography}
[{\includegraphics[width=1in,height=1.25in,clip,keepaspectratio]{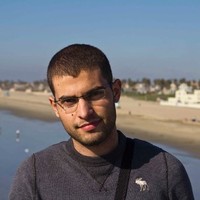}}]{Samet E. Arda} received his M.S. and Ph.D. degrees in Electrical Engineering from Arizona State University in 2013 and 2016. He is currently an Assistant Research Scientist at ASU. His Ph.D. thesis focused on development of dynamic models for small modular reactors. His current research interests include system-level design and  optimization techniques for scheduling in heterogeneous SoCs.
\end{IEEEbiography}
\vskip -1.0 \baselineskip plus -1fil
\begin{IEEEbiography}
[{\includegraphics[width=1in,height=1.25in,clip,keepaspectratio]{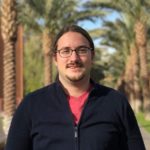}}]{A. Alper Goksoy} received his B.S. degree in Electrical and Electronics Engineering from Bogazici University, Istanbul, Turkey.  He is currently pursuing his Ph.D. in Electrical Engineering at Arizona State University.
\end{IEEEbiography}
\vskip -1.0 \baselineskip plus -1fil
\begin{IEEEbiography}
[{\includegraphics[width=1in,height=1.25in,clip,keepaspectratio]{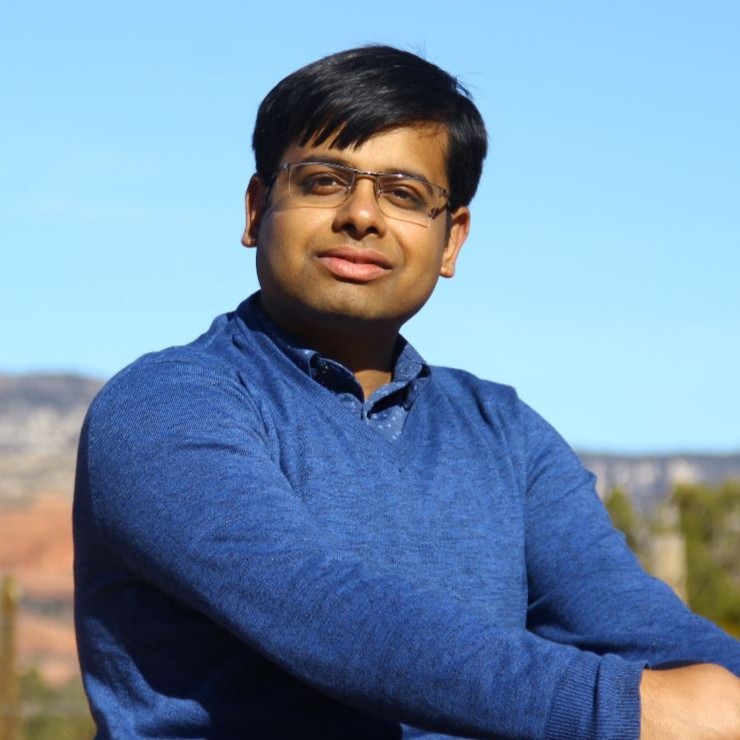}}]{Sumit K. Mandal} received his dual degree (BS + MS) from Indian Institute of Technology (IIT), Kha- ragpur. Currently, he is pursuing his Ph.D. in Arizona State University. His research interest includes analysis and design of NoC architecture, power management of multicore processors and AI hardware.
\end{IEEEbiography}
\vskip -1.0 \baselineskip plus -1fil
\begin{IEEEbiography}
[{\includegraphics[width=1in,height=1.25in,clip,keepaspectratio]{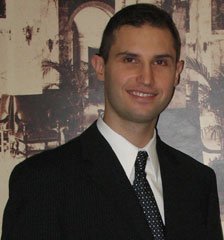}}]{Umit Y. Ogras} received his Ph.D. degree in Electrical and Computer Engineering from Carnegie Mellon University, Pittsburgh, PA, in 2007. From 2008 to 2013, he worked as a research scientist at the Strategic CAD Laboratories, Intel Corporation. He is an Associate Professor at the School of Electrical, Computer and Energy Engineering, and the Associate Director of WISCA Center.
\end{IEEEbiography}
\vskip -1.0 \baselineskip plus -1fil
\begin{IEEEbiography}
[{\includegraphics[width=1in,height=1.25in,clip,keepaspectratio]{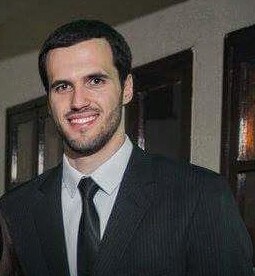}}]{Anderson L. Sartor} received his B.Sc. and Ph.D. in Computer Engineering from Universidade Federal do Rio Grande do Sul (UFRGS), Brazil, in 2013 and 2018, respectively. He is currently a Postdoctoral Researcher at Carnegie Mellon University (CMU). His primary research interests include embedded systems design, machine learning for energy optimization, SoC modeling, and adaptive processors.
\end{IEEEbiography}
\vskip -1.0 \baselineskip plus -1fil
\begin{IEEEbiography}
[{\includegraphics[width=1in,height=1.25in,clip,keepaspectratio]{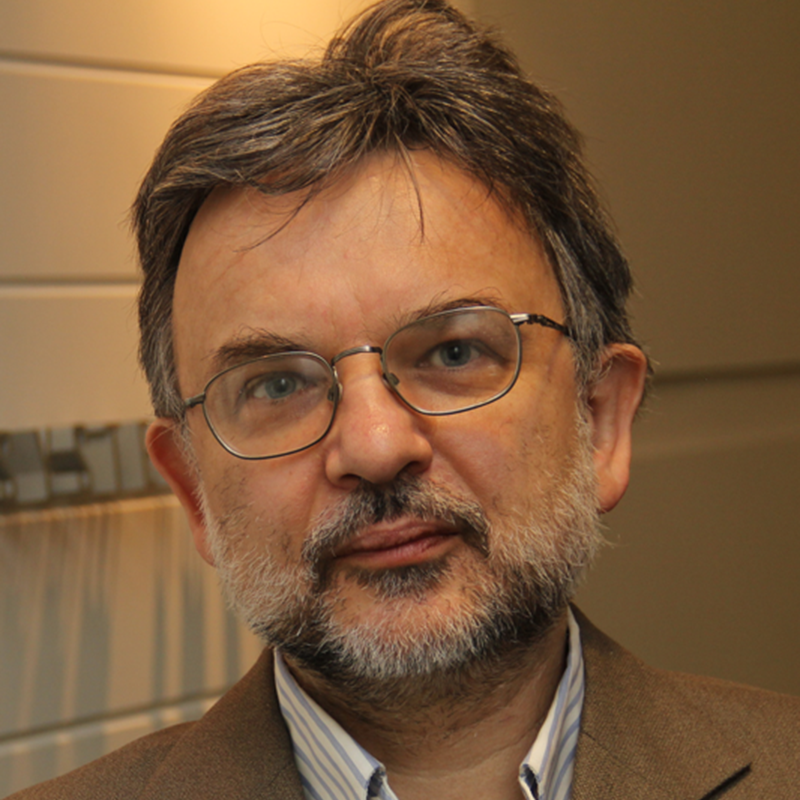}}]{Radu Marculescu} is the Laura Jennings Turner Chair in Engineering and Professor in the Electrical and Computer Engineering department at The University of Texas at Austin. He received his
Ph.D. in Electrical Engineering from the University of Southern California in 1998. Radu’s current research focuses on developing ML/AI methods and tools for modeling and optimization of embedded
systems, cyber-physical systems, and social networks.
\end{IEEEbiography}

\end{document}